\documentclass[fleqn,twoside]{article}
\usepackage{espcrc2}

\usepackage{graphicx}

\newcommand{\AmS}{{\protect\the\textfont2
  A\kern-.1667em\lower.5ex\hbox{M}\kern-.125emS}}

\title{Vacuum Structure of the Ichimatsu-Decomposed Lattice Models}

\author{K. Itoh\address{Faculty of Education, 
        Niigata University, Niigata 950-2181, Japan},
        M. Kato\address{Institute of Physics, University of Tokyo, Komaba, 
                        Meguroku, Tokyo 153, Japan},
        M. Murata\address[DPNU]{Department of Physics, Niigata University, 
        Niigata 950-2181, Japan},
        H. Sawanaka\addressmark[DPNU]
        and
        H. So\addressmark[DPNU]\thanks{Talk presented by H. So. This work was 
        supported in part by Grants-in-Aid for Scientific Research No. 12640259
        and 13135209 from the Japan Society for the Promotion of Science.}
       }
\begin{document}

\begin{abstract}
We proposed an `Ichimatsu'-decomposed lattice gauge theory with
fermionic symmetries. The vacuum structures of the gauge sector with two
coupling constants ($\beta_p,\beta_c$) are investigated for
3-dimensional Z$_2$ and 4-dimensional SU(2) cases using mean-field
approximation and numerical simulation. We found two phases on the 
($\beta_p,\beta_c$) phase diagram for 3-dim. Z$_2$ case, 
while the diagram is a single phase for the latter.

\vspace{1pc}
\end{abstract}

\maketitle

\section{MOTIVATION}

The lattice realization of super Yang-Mills theory has been
considered very difficult though highly desirable as a non-perturbative
formulation\cite{c-v,k-s}. In our previous work and talk
\cite{ichi,sawa}, we reported the presence of exact fermionic symmetries
on the `Ichimatsu' lattice, a novel lattice.  If the symmetries survive
in the continuum limit, they would be related to a supersymmetry if not
to the BRS symmetry.

From the equation presented shortly, we know that the fermionic
symmetries cannot survive on the ordinary lattice.  Therefore it is of
crucial importance to study the continuum limit on the `Ichimatsu'
lattice.  
\section{CELL+PIPE MODEL}

Plaquettes on the regular lattice gauge theory are decomposed into two sets:

(i) plaquettes on cell\\
~~~~~~~~~~~~~ $U_{n,\mu\nu}$ on $\{ n_{\mu}+n_{\nu} \equiv  0 ~~(mod ~2) \} $;

(ii) plaquettes on pipe\\
~~~~~~~~~~~~~ $U_{n,\mu\nu}$ on $\{ n_{\mu}+n_{\nu} \equiv  1 ~~(mod ~2) \} $.

One a $(\mu,\nu)$ plane, one finds an alternating or chequered
 pattern and it is called as an `Ichimatsu' pattern
 \cite{ichi,sawa}. The fields of the Ichimatsu lattice theory are link
 variables, $U_{n,\mu}$, as the gauge field, and a real staggered fermion,
 $\xi_n$, as the fermion field.

The gauge action of our model (cell+pipe) is
$$
S_{\rm gauge} = - \frac{\beta_c}{2} \sum_{\rm  ~on~ cell} {\rm Tr~}U_{n,\mu\nu} 
 - \frac{\beta_p}{2} \sum_{\rm  ~on~ pipe} {\rm Tr~}U_{n,\mu\nu} .
$$

The fermionic symmetry transformation of the fermion field is written as 
$$
\delta \xi_n \sim (1-\frac{\beta_p}{\beta_c}) \sum_{0<\mu<\nu}
(-)^{n_{\mu}+n_{\nu}}(C^{(+)}_{n,\mu\nu} + C^{(-)}_{n,\mu\nu})F_{n,\mu\nu}
$$
\noindent
where $C^{\pm}_{n,\mu\nu}$ are  Grassman-odd parameters. 
 
It should be emphasized that on the usual regular lattice ($\beta_c
= \beta_p$), the transformation goes down to $O(a)$. Therefore it is
important to study the dynamical properties on the two-couplings space
($\beta_p,\beta_c$).  From now on, we switch off fermion effects in our
model and concentrate on the vacuum structure of the gauge sector.

\section{EXPECTED PROPERTIES OF OUR MODEL}

\subsection{Mean Field Approximation}

We take   a link variable $<U_{n,\mu}> = v_{\mu}$  as a mean field. 
This approximation is  not  gauge-invariant 
and it is  useful only in the  first view of the phase diagram. 

The results are  

(1) There exists an exact duality under $(\beta_c, \beta_p)  \leftrightarrow  (\beta_p, \beta_c)$, 
 that is, the partition functions are invariant under the interchange of couplings; 
$$
Z_{\rm MF}(\beta_c, \beta_p) = Z_{\rm MF}(\beta_p, \beta_c). 
$$

(2) In the cell model (defined by putting $\beta_p=0$) and  the pipe
model ($\beta_c=0$), phase transitions occur.

These results are independent of the spatial dimension and 
the gauge group. 
A rough sketch of the diagram is shown in Fig.~1.

\begin{center}
\includegraphics[width=6cm]{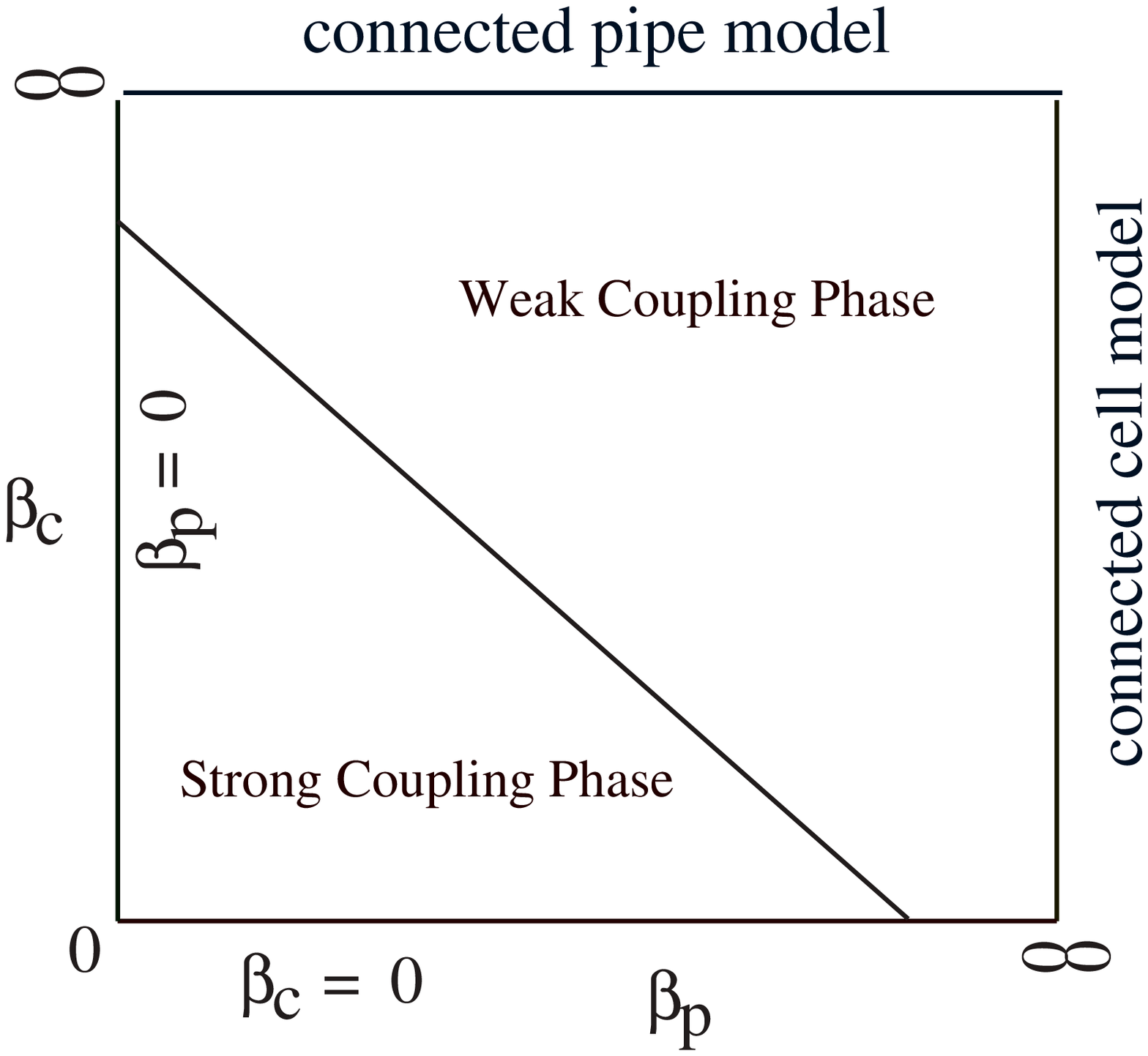} \\
\end{center}
\begin{small}
Fig.~1. ~Phase diagram of cell+pipe model in mean field approximation.
\end{small}


\subsection{Analytical Results}

1.~~ Regular lattice gauge theory ($\beta_c=\beta_p=\beta$)
 
When $\beta_p= \beta_c = \beta$, our model becomes a usual regular
lattice gauge theory. On the regular lattice, the 3D $Z_2$ gauge theory
has a phase transition point while the 4D SU(2) gauge theory has a
crossover point.

2.~~ Four phase boundaries

In order to understand the phase diagram, it is necessary to analyse
the limiting cases, $\beta_{c}~ {\rm or}~\beta_p \rightarrow 0 ~ {\rm
or}~ \infty$.  We can obtain analytical results for these limiting
cases.  In the cell model, $\beta_p=0$, the system is decomposed to a
finite degrees of freedom and it is exactly solvable. The pipe model,
$\beta_c=0$, is equivalent to one plaquette model and is also exactly
solvable. Both cases have no phase transition contrary to the result
of the mean field approximation.

In $\beta_{p}~ {\rm or}~\beta_c \rightarrow \infty$, plaquettes on cell
 or pipe are strongly connected. The dynamics for these situations is not known and
 the numerical analysis is necessary.

3.~~ Possibility of the continuum limit and its  phase transition

To take the continuum limit of our model, we must avoid the phase 
transition point or line. If we go  across the singularity, then we 
 miss the information on the vacuum, such as confinement. If one want to 
be consistent with confinement and asymptotic freedom, the phase diagram 
must be smoothly connected between the weak and strong
 coulping regions.

\section{PHASE DIAGRAM FOR ICHIMATSU PATTERN LATTICE}

\subsection{3D Z$_2$ Case}

The action is written as 
$$
 S = - \beta_c \sum_{\rm plaq. on~ cell} UUUU -   \beta_p \sum_{\rm plaq. on~ pipe} UUUU
$$
\noindent
where $U$ is a link variable taking the values, $\pm 1$.  Note the
normalisation of the coupling constants differs from the SU(2) case. 
For the regular lattice ($\beta_c=\beta_p = \beta$),  
it is found that there occurs  the 2nd order phase transition at $\beta=0.76$  
 obtained by  the analysis of the equivalent Ising spin. 
We have simulated 3D Z$_2$ cell+pipe model for the size, $4^3, 8^3$ and $16^3$. 
\begin{center}
\includegraphics[width=6.5cm]{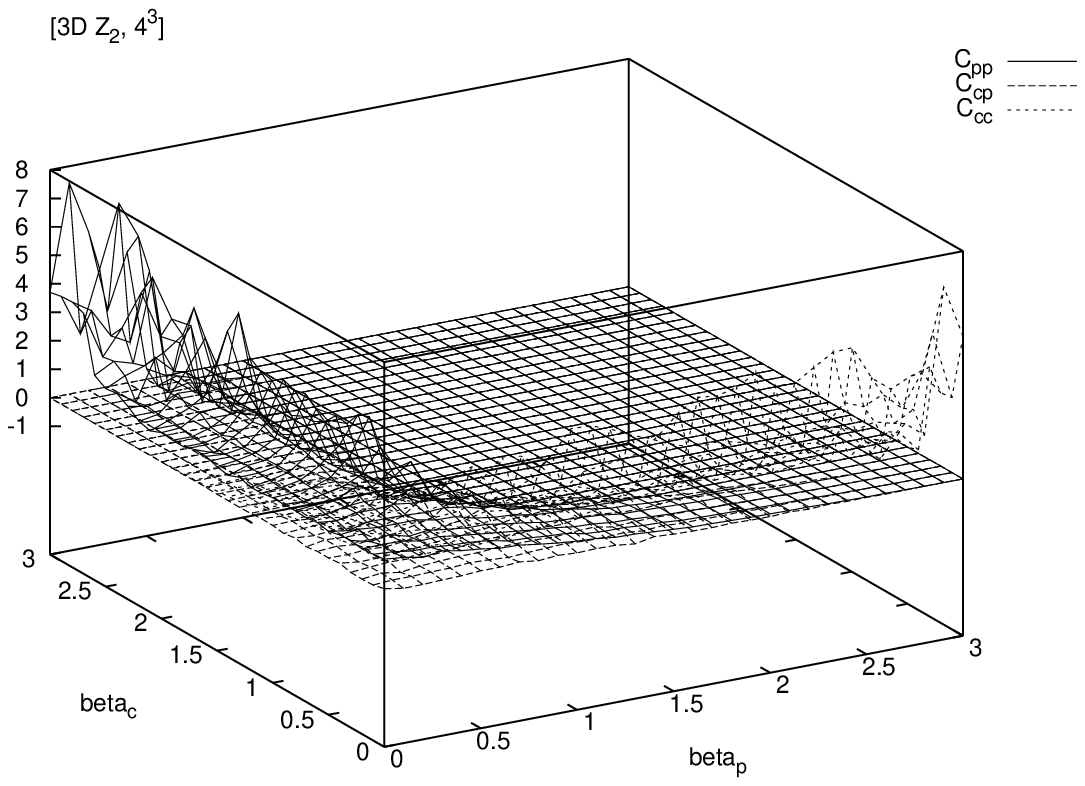} \\
\end{center}
\begin{small}
\vspace{-0.8 cm}
Fig.~2. 3D plot for ($\beta_p, \beta_c$, specific heat) in 3D Z$_2$model.
\end{small}

\vspace{0.3 cm}
From  Fig.~2, we can see that the phase transition point is
extended to the connected pipe ($\beta_c \rightarrow \infty, \beta_p=0.55$) model 
and  the connected cell ($\beta_p \rightarrow \infty, \beta_c=0.38$) model. 
\newpage
The phase diagram  for 3D Z$_2$ model is obtained in Fig.~3.
\begin{center}
\includegraphics[width=6cm]{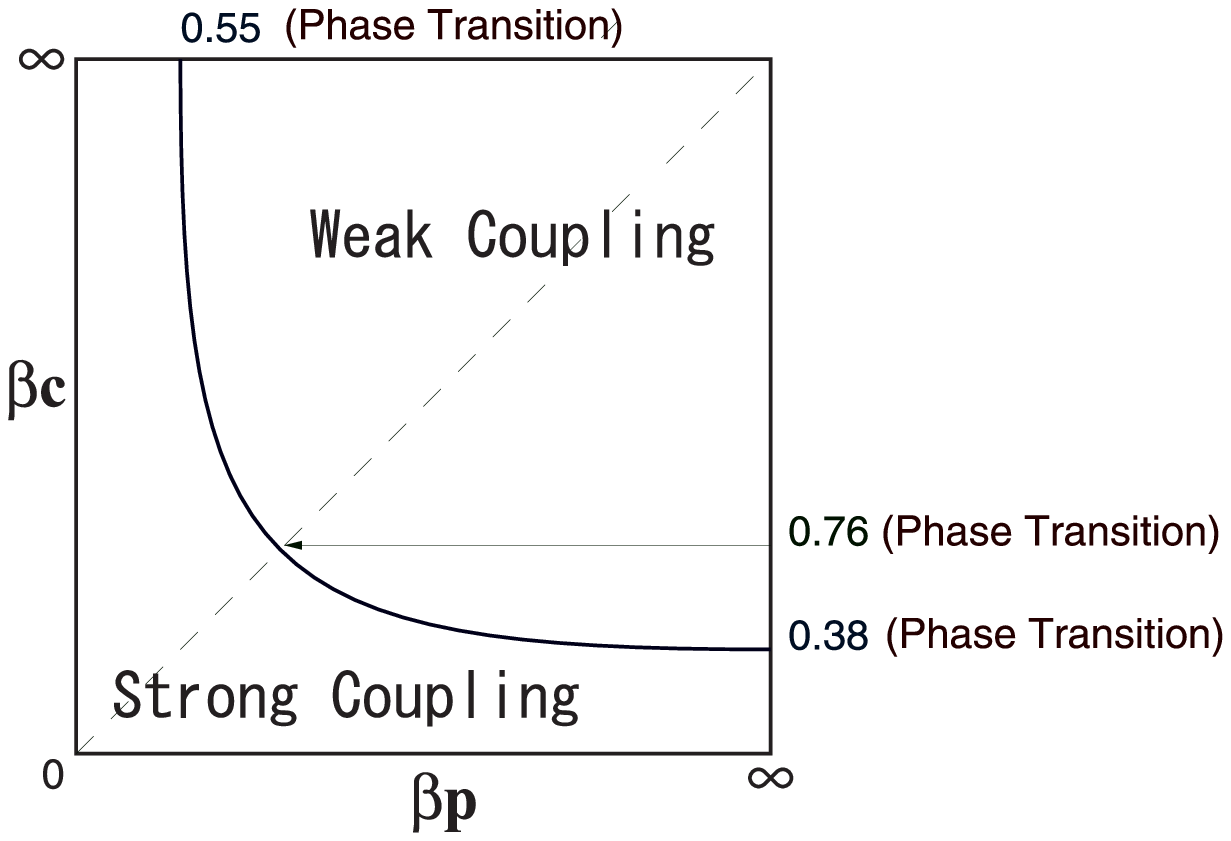} \\
\end{center}
\begin{small}
\vspace{-0.3 cm}
Fig.~3. Phase diagram for 3D Z$_2$ model.
\end{small}

\subsection{4D SU(2) case}

For lattice size, $8^4,12^4,16^4$, we  calculated the internal energy and the specific heat. 
In Fig.~4a, we show  a crossover behaviour ($\beta_p=\beta_c\sim 2.0$)  on a regular lattice and 
 a sharp peak of the specific heat for the connected cell model in Fig.~4b.
%
\begin{center}
\includegraphics[width=5cm]{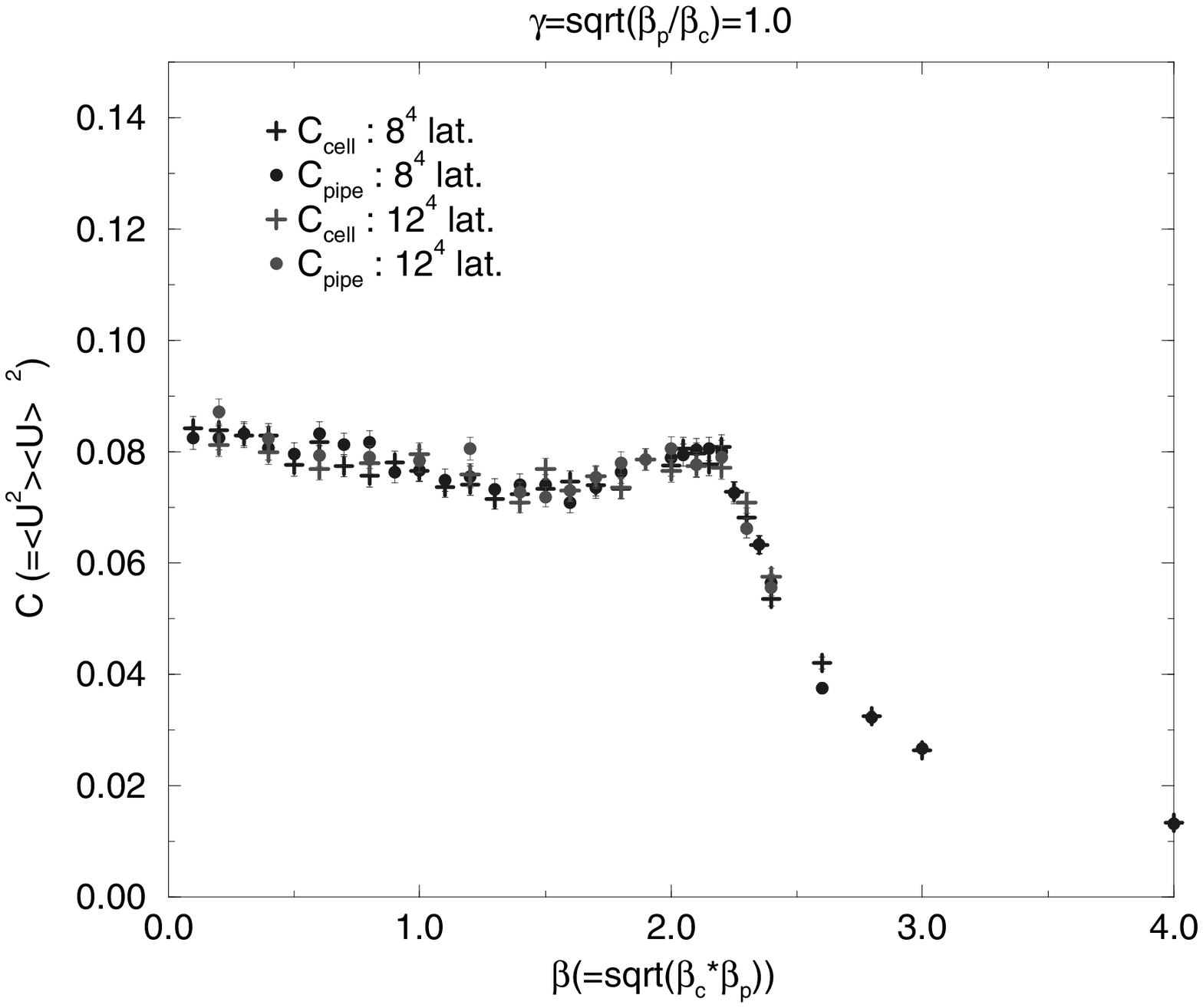} \\
\end{center}
\begin{small}
Fig.~4a. A crossover in the regular lattice model.
\end{small}
\begin{center}
\includegraphics[width=5cm]{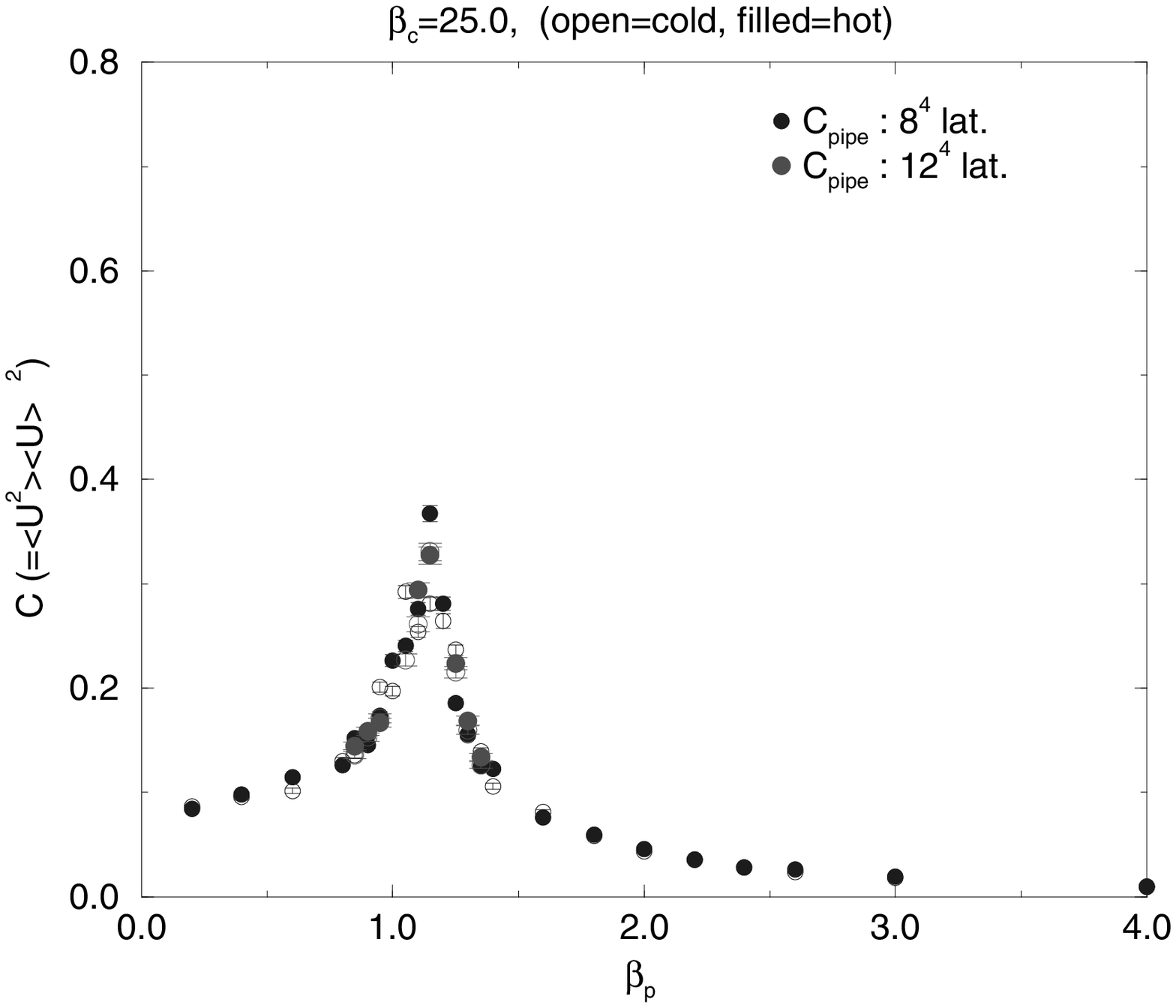} \\
\end{center}
\begin{small}
\vspace{-0.3 cm}
Fig.~4b. A peak of specific heat at $\beta_c=25$.
\end{small}
\vspace{0.3 cm}

Our simulations are consistent with the following statements: 
1) the peak in Fig.~4b is that of the 2nd order phase transition; 2) no
peak is present in the connected cell model; 3) a crossover line extends to
$(\beta_p \rightarrow \infty, \beta_c \sim 0)$.


\section{SUMMARY AND DISCUSSIONS}

To consider exact fermionic symmetries on lattice, we decomposed
all plaquettes into two sets, the cell and pipe. The vacuum structure of the gauge
sector was investigated to study its continuum limit.

Phase diagrams of 3D Z$_{2}$ and 4D SU(2) models were studied.  We found a
`corridor' (singularity free region) between strong coupling and weak
coupling regions for 4D SU(2) case (Fig.~5); the diagram is a single phase. The 3D Z$_2$ model
has two phases.

\begin{center}
\includegraphics[width=5.5cm]{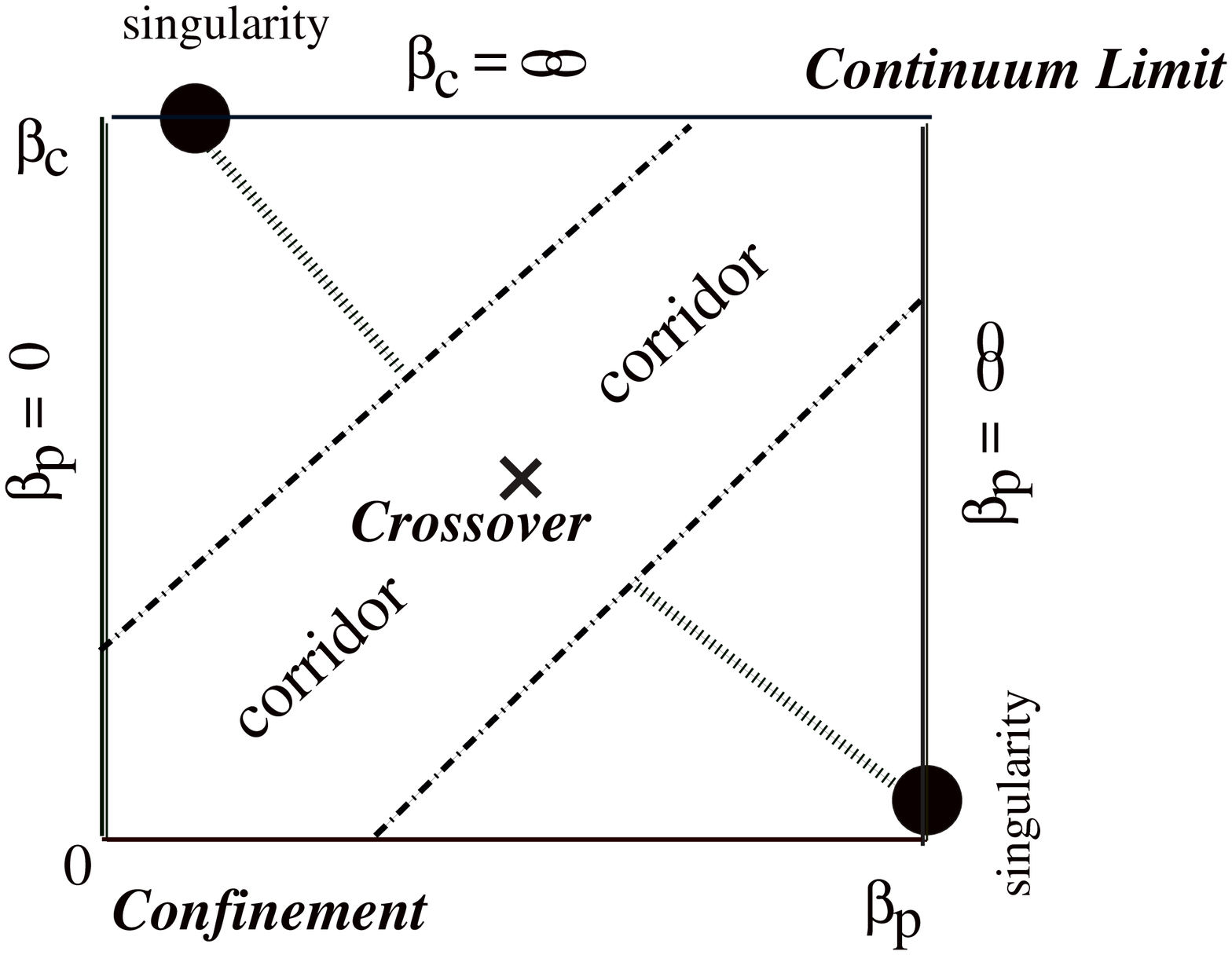}
\end{center}
\begin{small}
\vspace{-0.2 cm}
Fig.~5. Phase diagram for 4D SU(2) model.
\end{small}
\vspace{0.4 cm}

This is our first non-perturbative result on an `Ichimatsu' lattice gauge theory.
Further results along this direction will be reported in the forthcoming 
paper\cite{ichi2}.


\end{document}